\documentclass[12pt,a4paper]{iopart}
\usepackage{iopams,graphicx,hyperref,setstack}
\usepackage[usenames,dvipsnames]{color}
\usepackage{mathrsfs}
\DeclareSymbolFontAlphabet{\mathrsfs}{rsfs}

\makeatletter
\newcommand*{\rom}[1]{\expandafter\@slowromancap\romannumeral #1@}
\makeatother

\newcommand{\hateq}{\overset{\wedge}{=}}

\renewcommand{\[}{\begin{equation}}
\renewcommand{\]}{\end{equation}}
\newcommand{\half}{\textstyle \frac{1}{2}}
\newcommand{\Lie}{\mathcal{L}}
\newcommand{\tfrac}{\textstyle \frac}
\newcommand{\scri}{\mathrsfs{I}^+}
\newcommand{\norms}[1]{\Vert #1 \Vert^2}
\newcommand{\normbs}[1]{\Vert #1 \Vert_B^2}
\newcommand{\norm}[1]{\Vert #1 \Vert}
\newcommand{\normb}[1]{\Vert #1 \Vert_B}
\newcommand{\const}{\mathrm{const}}
\renewcommand{\Re}{\mathrm{Re}\,}

\begin{document}

\title{Numerical and analytical methods for asymptotically flat spacetimes}
\author{Oliver Rinne}
\address{Max Planck Institute for Gravitational Physics 
  (Albert Einstein Institute),\\Am M\"uhlenberg 1, 14476 Potsdam, Germany
  \\Department of Mathematics and Computer Science, Freie Universit\"at Berlin,
   Arnimallee 2--6, 14195 Berlin, Germany \smallskip \\
  E-mail: \texttt{oliver.rinne@aei.mpg.de}
}


\begin{abstract}
  This article begins with a brief introduction to numerical relativity aimed 
  at readers who have a background in applied mathematics but not necessarily 
  in general relativity.
  I then introduce and summarise my work on the problem of treating
  asymptotically flat spacetimes of infinite extent with finite computational
  resources.
  Two different approaches are considered.
  The first approach is the standard one and is based on evolution on Cauchy 
  hypersurfaces with artificial timelike boundary.
  The well posedness of a set of constraint-preserving
  boundary conditions for the Einstein equations in generalised harmonic gauge 
  is analysed, their numerical performance is compared with various alternate 
  methods, and improved absorbing boundary conditions are constructed and
  implemented.
  In the second approach, one solves the Einstein equations on hyperboloidal 
  (asymptotically characteristic) hypersurfaces.
  These are conformally compactified towards future null infinity,
  where gravitational radiation is defined in an unambiguous way.
  We show how the formally singular terms arising in a $3+1$ reduction of the
  equations can be evaluated at future null infinity, present stable numerical 
  evolutions of vacuum 
  axisymmetric black hole spacetimes and study late-time power-law tails of 
  matter fields in spherical symmetry.
\end{abstract}


\noindent
{\it Submitted as the introductory chapter of a Habilitation thesis 
consisting of the published papers \cite{Rinne2006}--\cite{Rinne2013}
to the Department of Mathematics and Computer Science at
Freie Universit\"at Berlin in November 2013}

\bigskip
\noindent
This article is organised as follows.
In section \ref{s:nr} we give an introduction to the basics of numerical
relativity, with a focus on the Cauchy problem, formulations of Einstein's 
equations and numerical methods.
In section \ref{s:outer} we introduce the main subject of this thesis,
the treatment of asymptotically flat spacetimes and the
``outer boundary problem'' in numerical relativity.
Section \ref{s:ibvp} summarises my work on the first approach to this
problem, namely Cauchy evolution with artificial timelike boundary.
Section \ref{s:scri} is devoted to a different approach based on hyperboloidal
evolution to future null infinity.
Finally in section \ref{s:concl} we conclude and give a brief outlook on 
future research directions.

\section{Numerical relativity}
\label{s:nr}

\subsection{A brief history}

Einstein's 1915 theory of general relativity has revolutionised the way
we think about gravitation.
Its radical difference from other field theories lies in the fact that its
equations govern the geometry of spacetime itself, as opposed to most other
theories where fields evolve on an unchanging background geometry.
The geometry of spacetime is determined by its matter content through 
Einstein's field equations. 
In turn, matter moves along geodesics of this spacetime manifold.
To put it simply, \emph{gravitation is geometry.}

Through observations such as the perihelion shift of Mercury, 
the bending of light in the gravitational field of the sun, 
the gravitational redshift, and the decrease of the orbital period of
binary pulsars consistent with the loss of energy due to emission of 
gravitational radiation (Hulse \& Taylor, Nobel prize 1993), general relativity 
is by now one of the most accurately verified physical theories.
Nevertheless, most of these observations only test the validity of the theory
in the weak-field limit.
Almost a century after Einstein's discovery, still relatively little is known
about the full implications of the theory in the nonlinear regime.

Aside from these astrophysical questions, there are several problems in
mathematical relativity that remain unanswered.
Two of the most important ones are the question of black hole stability and
the cosmic censorship conjecture.
Even though widely expected to be true, it was only in 1993 that Christodoulou
and Klainerman were able to prove in a voluminous work \cite{Christodoulou1993}
that flat (Minkowski) spacetime is nonlinearly stable.
Despite some recent progress, a similar theorem for the general stationary 
vacuum black hole, the Kerr solution, is still lacking.
This is of central importance as black holes are believed to be ubiquitous 
in the universe.

A different conjecture, first put forward by Penrose in 1969 \cite{Penrose1969}
and termed \emph{cosmic censorship}, concerns the global behaviour of solutions.
The Einstein equations are known to form singularities from quite general 
initial data \cite{HawkingEllis}.
The (weak) cosmic censorship conjecture states that (very roughly) any 
singularities formed from generic initial data lie inside an event horizon,
i.e.~they are causally disconnected from (invisible to) far-away observers.
So far there is no general proof of this conjecture, which has important 
consequences on the determinism of the theory.

Why then do Einstein's equations pose such tremendous difficulties to the
mathematician? 
Despite their elegant geometric origin, they turn out
to be a complicated system of coupled nonlinear second-order partial
differential equations (PDEs).
Exact solutions are generally only known under strong simplifying assumptions
such as the existence of spacetime symmetries.
Small perturbations of known solutions can be studied by linearising
the field equations.

One approach to studying the behaviour of more general solutions is the use
of numerical approximations.
Due to the complexity of the equations involved, this requires powerful
computers, and as a result \emph{numerical relativity} is a relatively
young field of research:
it started around 1964 with pioneering work by Hahn \& 
Lindquist \cite{Hahn1964}, who studied the head-on collision of two black 
holes.
Since then the field has had a history of several breakthroughs as well as long
periods of struggle. (Excellent recent textbooks on the subject are for example
\cite{Baumgarte2010,Alcubierre2008}.) 

Arguably one of the most important achievements made through numerical 
simulations is the discovery of critical phenomena in gravitational collapse
by Choptuik in 1993 \cite{Choptuik1993}.
This was triggered by a question posed by a mathematical relativist
(Christodoulou): 
consider a family of initial data corresponding to compact matter configurations
with one parameter, such that for small values of the parameter the
configuration will disperse to leave flat spacetime behind, whereas 
for large values it will collapse to form a black hole.
What happens at the threshold between the two outcomes?
Choptuik investigated this using sophisticated numerical methods
(most importantly, adaptive mesh refinement) and observed phenomena 
similar to thermodynamic phase transitions, including power-law scaling
of the black hole mass in supercritical evolutions and a universal,
self-similar critical solution.

The majority of researchers in numerical relativity focused on what was 
regarded as the most important outstanding problem in numerical relativity, 
the collision of two orbiting black holes.
Black holes being the simplest objects in general relativity, this is the
obvious analogue of the two-body problem in Newtonian gravity.
The problem received so much attention because binary black hole collisions
are widely considered to be the strongest sources of gravitational waves,
which are hoped to be detected directly in the near future by several
earth-based detectors already in operation, a planned space-based detector
(eLISA) that has just been approved by the European Space Agency 
to be launched in 2034, and alternative observational methods such as pulsar 
timing arrays.
There is thus a strong need for models of gravitational waveforms from
astrophysical events to be used for matched filtering in gravitational 
wave data analysis.
Despite much effort spent on the binary black hole problem,
it was not until 2005 that the final breakthrough was made and the first
complete simulations of the inspiral, merger and ringdown of a black hole
binary were presented almost simultaneously by three different groups
\cite{Pretorius2005a,Campanelli2006,Baker2006}.
By now such simulations have almost become routine.
Wider regions of the parameter space have been explored, matter has been
included (binary neutron stars or neutron star/black hole binaries) and more
complicated physics is being added.
These ``numerical laboratories'' serve as substitutes for experiments on 
astronomical scales---an interesting philosophical shift of paradigm.


\subsection{The Cauchy problem for the Einstein equations}

In order to understand why the numerical solution of Einstein's equations poses
such difficulties, let us consider the general structure of these equations.
Spacetime is described by a smooth four-dimensional manifold $M$ with a smooth
Lorentzian metric $g_{ab}$.\footnote{Throughout we use abstract index notation,
  whereby $g_{ab}$ represents the $0 \choose 2$ tensor field $g$ on $M$.
  Indices $a,b, \ldots$ range over $0,1,2,3$. 
  The notation in this chapter has been streamlined to be self-consistent; 
  it differs from the notation used in some of the following chapters.}
The Einstein equations are
\begin{equation}
  \label{e:einstein}
  G_{ab} = \kappa T_{ab}.
\end{equation}
Here $G_{ab} = R_{ab} - \half R g_{ab}$ is the Einstein tensor, $R_{ab}$ is 
the Ricci tensor and $R$ the scalar curvature.
These are evaluated with respect to the Levi-Civita connection compatible
with $g_{ab}$.
On the right-hand side, $T_{ab}$ is the energy-momentum tensor describing the 
matter content of spacetime, and $\kappa$ is a constant.
For the time being we may assume vacuum, $T_{ab} = 0$.
Equation \eref{e:einstein} is to be solved for the metric $g_{ab}$;
it thus forms a system of second-order, quasi-linear PDEs.
A key property of \eref{e:einstein} is its invariance under arbitrary smooth
transformations of the spacetime coordinates $x^a$, a principle often referred 
to as general covariance.

However, in order to solve the equations numerically, one needs to pick a 
particular coordinate chart in order to obtain a definite set of PDEs.
This is most often done using the Cauchy or initial-value formulation of 
general 
relativity.\footnote{A different approach is the characteristic formulation; 
see \cite{WinicourLRR} for a review and also section \ref{s:matching}
in this chapter.}
For this one picks a time coordinate $t := x^0$ and considers a foliation 
of spacetime into the slices $\Sigma(t)$ of constant time $t$.
Indices $i,j,\ldots$ from the middle of the alphabet will be used to denote the
spatial coordinates $x^i$, $i=1,2,3$.
The Einstein equations \eref{e:einstein} split into two different classes.
The equations for which both indices are spatial ($ab=ij$) are found to 
contain second time derivatives of the metric; these six equations are 
called \emph{evolution equations}.
The equations for which one index is temporal (say $a=0$) are found to contain
no second time derivatives of the metric; these four equations are therefore
called \emph{constraint equations}.
The constraint equations are preserved under the time evolution in the sense
that if the constraints vanish at one instant of time then the evolution
equations imply that their time derivatives vanish as well.
This is a consequence of the contracted Bianchi identities
\[
  \nabla^b G_{ab} = 0,
\]
where $\nabla$ denotes the covariant derivative compatible with $g_{ab}$.
While this is true on the analytical level, numerical simulations have long
been plagued by exponentially growing constraint violations.
Only relatively recently has this problem been cured (see below in 
section \ref{s:gh}).

On an initial spacelike hypersurface $\Sigma_0$ corresponding to $t=0$,
we specify initial data for $g_{ab}$ and $\partial_t g_{ab}$ satisfying
the constraint equations.
(Constructing such data is itself a highly nontrivial problem, see 
\cite{CookLRR} for a review.)
The evolution equations are then integrated forward in time in order to obtain
$g_{ab}$ for $t>0$.
There is a slight problem though: we have ten unknowns $g_{ab}$ but only
six evolution equations.
At this point general covariance comes into play: fixing the coordinates 
allows us to impose four conditions on the components of $g_{ab}$, the
so-called coordinate or \emph{gauge} conditions.
Thus we really only have six free components of the metric that are evolved
by the six evolution equations.

In a numerical simulation it is difficult if not impossible to fix the
spacetime coordinates \emph{a priori} as one does not usually know what
spacetime a given set of initial data will evolve to.
Instead one ties the coordinates to the dynamical fields, hoping that the
coordinates that are thus being constructed ``on the fly'' will have desirable
properties (e.g., avoidance of singularities).
Depending on how this is done, the final set of PDEs one obtains may take on
very different forms.
In fact, the Cauchy problem may be well posed or ill posed!
In the following subsection we briefly review the two formulations of the
Einstein equations that are most often used in numerical relativity and,
in fact, in the present thesis.


\subsection{Formulations of the Einstein equations}
\label{s:formulations}

\subsubsection{Generalised harmonic coordinates}
\label{s:gh}

One way to fix the spacetime coordinates is to impose a wave equation 
on each of the coordinates $x^a$:\footnote{Note the d'Alembert operator is
meant to act on each of the coordinates as scalar functions here.}
\[
  \label{e:ghgauge}
  \Box x^a \equiv g^{bc} \nabla_b \nabla_c (x^a) = -g^{bc} \Gamma^a{}_{bc} = H^a,
\]
where $\Gamma^a{}_{bc}$ denotes the Christoffel symbols of the Levi-Civita 
connection.
Such coordinates are called \emph{(generalised) harmonic}.
The source functions $H^a$ on the right-hand side may depend on the 
coordinates $x^a$ and on the metric $g_{ab}$ but not on derivatives of the 
metric.

With this gauge condition the vacuum Einstein equations can be written as
\[
  \label{e:gheinstein}
        g^{cd} \partial_c \partial_d \, g_{ab} = 
        - \nabla_{a} H_{b} - \nabla_{b} H_{a} + 2 g^{cd} g^{ef} (\partial_e g_{ca} 
        \partial_f g_{db} - \Gamma_{ace} \Gamma_{bdf}),
\]
i.e. the principal part of the equation becomes the d'Alembert operator
associated with the metric.
Hence the system of PDEs is symmetric hyperbolic, a fact that was used
by Four\`es-Bruhat in her celebrated proof of the well-posedness of the Cauchy
problem for the Einstein equations \cite{FouresBruhat1952}.

Yet it was only much later that harmonic coordinates made their way into
numerical relativity.
Pretorius' 2005 breakthrough binary black hole 
simulations \cite{Pretorius2005a} were based on this system.

A crucial ingredient was a new method to control the growth of constraint 
violations.
In the generalised harmonic formulation, the role of the constraints is
taken on by the quantities
\[
  \mathcal{C}^a := g^{bc} \Gamma^a{}_{bc} + H^a,
\]
which must vanish for a solution to the Einstein equations because of the
gauge condition \eref{e:ghgauge}.
The evolution equation \eref{e:gheinstein} implies the following evolution
equation for the constraints:
\[
  \label{e:ghconstrevol}
  \nabla^b \nabla_b \mathcal{C}_a + \mathcal{C}^b \nabla_{(a} \mathcal{C}_{b)} 
  = 0.
\]
A linear stability analysis of this equation shows that not all 
modes decay, and they may be amplified due to the nonlinearity of the equation.
The key idea now is that we are still free to add multiples of the constraints
$\mathcal{C}_a$ to \eref{e:gheinstein} because these vanish for a solution
to Einstein's equations. 
Such terms will not affect the principal part of \eref{e:gheinstein} because the
constraints contain only first derivatives of the metric.
Adding constraints to \eref{e:gheinstein} will modify the constraint evolution
equation \eref{e:ghconstrevol}.
In \cite{Gundlach2005} a particular combination of such \emph{constraint damping
terms} was devised such that on the linear level all non-constant modes of the 
modified constraint evolution equation decay.

The generalised harmonic formulation of the Einstein equations forms the basis
of the first part of this thesis (chapters II--IV).
More precisely, we use a first-order reduction (with respect to time and 
spatial derivatives) of \eref{e:gheinstein}
developed by the Caltech-Cornell numerical relativity collaboration.
Details of this reduction can be found in \cite{Lindblom2006} and in 
\cite{Rinne2006}.

\subsubsection{ADM formulation}
\label{s:adm}

\begin{figure}
  \centering
  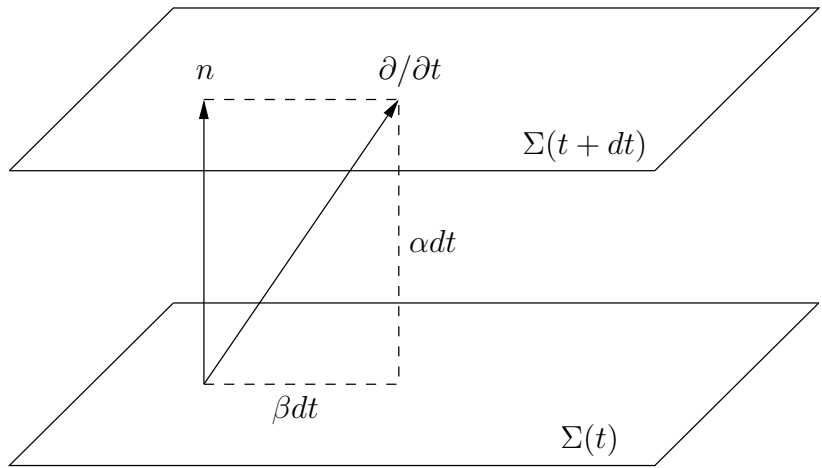
  \caption{\label{f:adm} 
  $3+1$ decomposition with unit timelike normal $n^a$, lapse function $\alpha$ 
  and shift vector $\beta^i$.
  }
\end{figure}

Before the introduction of generalised harmonic coordinates in numerical 
relativity, most numerical work was based on the $3+1$ or \emph{ADM formulation}
of the Einstein equations originally developed by Arnowitt, Deser and Misner 
in 1962 with a view towards quantising gravity (\cite{Arnowitt1962}; see 
also \cite{York1979}).
In this framework one decomposes the vector field $\partial/\partial t$ 
associated with the time coordinate $t$ into a part normal to the hypersurface
$\Sigma(t)$ of constant $t$ and a part tangential to it:
\[
  \left(  \frac{\partial}{\partial t} \right)^a 
  = \beta^a + \alpha n^a,
\]
where $n^a$ denotes the unit timelike normal to $\Sigma(t)$, $\alpha$ is the 
lapse function and $\beta^i$ the shift vector\footnote{Since $\beta^a$ is 
tangential to $\Sigma(t)$, it has only three nonvanishing components, 
hence we write it as $\beta^i$.}(figure \ref{f:adm}).

The spacetime metric takes the form
\[
  g = -\alpha^2 \rmd t^2 + \gamma_{ij}(\rmd x^i + \beta^i \rmd t)(\rmd x^j 
  + \beta^j \rmd t),
\]
where $\gamma_{ij}$ is the spatial metric (first fundamental form) induced on 
$\Sigma(t)$.
We also need to introduce the extrinsic curvature (second fundamental form)
\[
  \label{e:lngammaij}
  K_{ij} =  -\half \Lie_n \gamma_{ij},
\]
where $\Lie$ denotes the Lie derivative,
$\Lie_n = \alpha^{-1} ( \partial_t - \Lie_\beta )$.
Equation \eref{e:lngammaij} can be regarded as an evolution equation for 
$\gamma_{ij}$.
The vacuum Einstein equations imply an evolution equation for $K_{ij}$,
\[
  \label{e:lnkij}
        \Lie_n K_{ij} = - \alpha^{-1} D_iD_j\alpha + \mathcal{R}_{ij} 
        - 2K_{ik}K^k{}_j + K_{ij} K,
\]
where $D$ denotes the covariant derivative compatible with 
$\gamma_{ij}$, $\mathcal{R}_{ij}$ is the Ricci tensor of $\gamma_{ij}$,
and $K = \gamma^{ij} K_{ij}$.
The constraint equations take the form
\begin{eqnarray}
        \label{e:hamcons}
        \mathcal{H} &:=& \mathcal{R} + K^2 - K_{ij} K^{ij} = 0,\\
        \label{e:momcons}
        \mathcal{M}^j &:=& D_i (K^{ij} - \gamma^{ij} K) = 0,
\end{eqnarray}
where $\mathcal{R}$ is the scalar curvature of $\gamma_{ij}$.

It was only realised in the numerical relativity community in the 1990s that
for fixed lapse and shift, the ADM evolution equations \eref{e:lngammaij} and
\eref{e:lnkij} are only weakly hyperbolic and hence the initial value problem
is ill posed (see \cite{ReulaLRR} for a review of hyperbolicity for the
Einstein equations).

One way to cure this is to add multiples of the constraints, especially the 
momentum constraint \eref{e:momcons}, to \eref{e:lnkij}.
This was the essential trick that led to the formulation of Baumgarte,
Shapiro, Shibata and Nakamura \emph{(BSSN)} \cite{Shibata1995,Baumgarte1998}, 
which in addition to the 
generalised harmonic formulation has become one of the two standard
formulations used in binary black hole simulations.

A different approach, taken in the second part of this thesis, is the use
of elliptic gauge conditions.
As a condition on the spacetime slicing we shall require the mean curvature
$K$ of the slices to be a spacetime constant.
Apart from its geometric appeal, this will furnish the desired asymptotic
behaviour of the slices (see section \ref{s:outer_scri}).
Such slices also have good singularity avoidance properties as the mean
curvature controls the time evolution of the spatial volume element 
$\sqrt{\det \gamma_{ij}}$.
Preservation of the constant mean curvature (CMC) condition under the time 
evolution leads to an elliptic equation for the lapse function $\alpha$.
The spatial coordinates will be required to be \emph{spatially} harmonic,
i.e.,
\[
  \label{e:shgauge}
  \Delta x^i \equiv \gamma^{jk} D_j D_k (x^i) = 
  -\gamma^{jk} \; {}^{(3)} \Gamma^i{}_{jk} = H^i,
\]
where the $H^i$ are fixed functions of the spatial coordinates
(cf.~\eref{e:ghgauge}; now ${}^{(3)} \Gamma^i{}_{jk}$ refers to the
Christoffel symbols of $\gamma_{ij}$).
Taking a time derivative of \eref{e:shgauge} results in an elliptic equation
for the shift vector $\beta^i$.
It has been shown at least in the spatially compact case that the ADM 
system with these elliptic gauge conditions (CMC slicing and spatially harmonic
gauge) has a well-posed initial value problem \cite{Andersson2003}.
The price to pay is that we need to solve elliptic equations at each time
step of the numerical evolution, which is generally more computationally 
expensive than solving hyperbolic equations.

As mentioned earlier, due to general covariance, there is a redundancy in 
Einstein's equations that allows one to solve only the evolution 
equations\footnote{The constraints always need to be solved at the initial 
time.}; the constraints will be preserved under the time evolution. 
(Of course one still needs to check that violations of the 
constraints remain small during a numerical evolution.)
This approach is referred to as \emph{free evolution}.
A different approach, which we shall adopt in the second part of this thesis,
is \emph{constrained evolution}, whereby the constraints \eref{e:hamcons}
and \eref{e:momcons} are solved explicitly in lieu of some of the evolution 
equations.
This will give us better control of the asymptotic behaviour of the fields;
constrained evolution schemes are also often found to be more stable in highly 
nonlinear gravitational collapse simulations.
Of course the constraints add to the number of elliptic equations 
to be solved at each time step.


\subsection{Numerical methods}

Once we have decided on a particular formulation of the Einstein equations, 
the question arises which numerical methods should be used to solve this
system of PDEs.
Here we briefly review the two methods that are most often used in
numerical relativity: pseudo-spectral methods and finite-difference methods.
These methods work well for smooth solutions, which is the case for the vacuum
Einstein equations and also for most radiative forms of matter (e.g., scalar, 
electromagnetic or Yang-Mills fields).
For matter that may form discontinuities, e.g.~perfect fluids, these methods
are generally not suitable.
In this case finite-volume methods are normally used for the matter evolution
equations.

\subsubsection{Pseudo-spectral methods}
\label{s:ps}

The basic idea of spectral methods is an expansion of the numerical 
approximation $u(x)$ in a known set of basis functions $u_n(x)$,
here in one dimension for simplicity:
\[
  \label{e:spectral}
  u(x) = \sum_{n=0}^N a_n u_n(x).
\]
The $u_n(x)$ usually belong to a complete orthonormal set of functions.
In the spherical topology that is most often encountered in numerical 
relativity,
one usually expands in Chebyshev polynomials in the radial direction and
spherical harmonics in the angular directions.
Hereby the radial direction is often divided into a few subdomains and an 
expansion of the form \eref{e:spectral} is used in each of the subdomains.

Derivatives of $u(x)$ can be computed \emph{exactly} within the approximation
\eref{e:spectral} using the known derivatives of the basis functions.
In order to compute nonlinear terms, \emph{pseudo-}spectral methods
evaluate the approximation $u(x)$ at a discrete set of collocation points
$x_i$, usually the Gauss- or Gauss-Lobatto points of the numerical quadrature
associated with the basis functions.
Nonlinear terms are evaluated at these collocation points and thereafter
the spectral expansion coefficients $a_n$ of the result are computed.

For smooth solutions, pseudo-spectral methods converge exponentially with the
number $N$ of expansion coefficients.
Hence $N$ is usually taken to be quite small, $N\lesssim 50$. 
For larger $N$ roundoff errors quickly spoil any further gain in accuracy.

\subsubsection{Finite-difference methods}
\label{s:fd}

Finite-difference methods are based on an expansion of the solution in
a (finite) Taylor series.
Derivatives are replaced with difference quotients, e.g. for a one-dimensional 
uniform grid with spacing $h$:
\[
  (u')_i = \frac{1}{2h} (u_{i+1} - u_{i-1}) + \Or(h^2),
\]
where $u_i := u(x_i)$.
Near a boundary, one-sided operators are often used, e.g. for a right boundary
at $x=x_N$:
\[
  (u')_N = \frac{1}{2h} (3 u_N - 4 u_{N-1} + u_{N-2}) + \Or(h^2).
\]
The above are examples of second-order accurate finite difference operators;
in the second part of this thesis we will work with fourth-order accurate
finite differences. 

A subtle point is the treatment of coordinate singularities, e.g. for
axisymmetric spacetimes on the axis of symmetry $\rho = 0$ in cylindrical
polar coordinates $\rho, z, \phi$.
For this we use a staggered grid, where the first grid point is at $x_1 = h/2$,
and we add a \emph{ghost point} at $x_0 = -h/2$. 
(One ghost point suffices for second-order accurate finite differences; 
two are needed for fourth-order accuracy.)
The evolved fields are either even or odd with respect to $\rho$.
For an even function $u$ we set $u_0 = u_1$, whereas for an odd function we set
$u_0 = - u_1$.
This allows us to use centred finite difference operators at all interior
points $i\geqslant 1$.

\subsubsection{Multigrid for elliptic equations}
\label{s:mg}

For the constrained evolution schemes considered in the second part of this
thesis, elliptic equations need to be solved at each time step and hence
an efficient elliptic solver is needed.
The matrices arising from finite-difference approximations to elliptic
equations are sparse.
Standard relaxation method such as Gauss-Seidel relaxation are efficient
in damping short-wavelength components of the numerical error.
The slow convergence for longer wavelengths can be accelerated by using
a hierarchy of grids with increasingly coarser grid spacings, between which
the numerical approximation is transferred:
the multigrid method (\cite{Brandt1977}; an excellent concise introduction
is \cite{BriggsMG}).
We use the Full Approximation Storage variant of the algorithm in order
to treat nonlinearities in the equations directly, combined with a 
nonlinear Gauss-Seidel relaxation.

\subsubsection{Time integration}
\label{s:mol}

A framework often used in numerical relativity is the method of lines:
the equations are first discretised in space and then regarded as a large
system of ordinary differential equations (ODEs) in time, one at each 
grid/collocation point.
Standard ODE methods (e.g.~Runge-Kutta) can be used to integrate these ODEs
forward in time.

Some care must be taken in order to insure stability of the method,
in addition to the usual Courant-Friedrichs-Lewy condition on the timestep.
Finite-difference methods typically require artificial 
Kreiss-Oliger \cite{Kreiss1989,Kreiss1995} dissipation for stability in the 
context of the method of lines. 
It is important to note though
that these extra terms are below the level of the truncation error.
Pseudo-spectral methods often suffer from aliasing arising from the pointwise
evaluation of nonlinear terms.
This can be cured by some form of spectral filtering \cite{Boyd2001}.
An example is Orszag's Two-Thirds rule, whereby the upper third of the
expansion coefficients is set to zero prior to evaluation of nonlinear terms.


\section{The outer boundary problem for isolated systems}
\label{s:outer}

A common task one faces in numerical relativity is the modelling of an
\emph{isolated system}, i.e.~a compact self-gravitating object, e.g.~a star,
surrounded by an asymptotically flat spacetime.
Here asymptotically flat means in a very loose sense that the spacetime metric
approaches the Minkowski metric in the limit of infinite distance from the
source.
It should be stressed that this picture is an idealisation: of course the
universe is full of compact objects, and whether the universe
is asymptotically flat is a matter of debate.
Nevertheless, if we are only interested in the dominant contribution of one
particular distant object to, say, the gravitational radiation observed on the 
earth, then it is often a good approximation to surround this object by 
an asymptotically flat vacuum spacetime and to consider ourselves to be at 
infinite distance from the source.
The problem then arises to model an asymptotically flat spacetime of infinite 
extent with finite computational resources, and this is the main subject of 
this thesis.


\subsection{Conformal infinity}
\label{s:confinf}

In order to illustrate the various approaches to this problem, it is convenient
to adopt Penrose's idea of \emph{conformal compactification} \cite{Penrose1965}.
We write the spacetime metric as a conformal factor times a conformally
related metric:
\[ 
  \label{e:confdecomp}
   g_{ab} = \Omega^{-2} \tilde g_{ab}.
\]
Now we map the spacetime coordinates to a compact region such that $\Omega$
vanishes at the boundary, and $\tilde g_{ab}$ is everywhere finite when evaluated
in components with respect to the compactified coordinates. 

As an example, consider Minkowski spacetime
\[
  g = -\rmd t^2 + \rmd r^2 + r^2 \sigma,
\]
where $\sigma := \rmd \theta^2 + \sin^2\theta\, \rmd \phi^2$ is the round 
metric on the unit sphere.
Performing the coordinate transformations
\[
  \fl u=t-r, \quad v=t+r, \quad p = \arctan u, \quad q = \arctan v, \quad 
  T = p+q, \quad R=q-p,
\]
the metric can be written in the form \eref{e:confdecomp} with
\[
  \label{e:confmink}
  \Omega = 2 \cos p \, \cos q, \qquad 
  \tilde g = -\rmd T^2 + \rmd R^2 + (\sin^2 R) \, \sigma.
\]
Hence Minkowski spacetime is conformally related to the manifold 
$\mathbb{R} \times S^3$ with standard metric.
However we obtain only part of this ``Einstein cylinder'':
the ranges of the compactified coordinates are
\[
  \fl -\tfrac{\pi}{2} < p \leqslant q < \tfrac{\pi}{2} 
  \quad \Rightarrow \quad
  -\pi < T < \pi, \quad 0 \leqslant R < \pi, \quad T+R < \pi, \quad T-R > -\pi.
\]

\begin{figure}
  \centering
  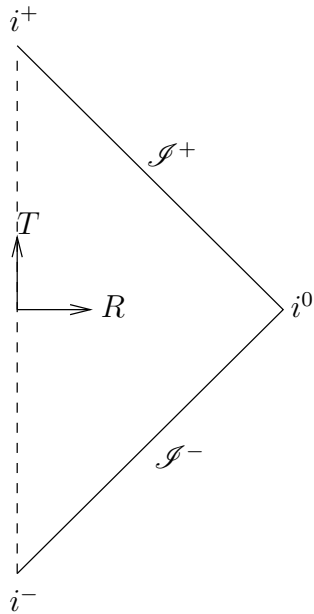
  \caption{\label{f:penrose} 
  Penrose diagram of Minkowski spacetime.
  }
\end{figure}

The resulting \emph{Penrose diagram} is shown in figure \ref{f:penrose}.
Since the mapping is conformal, light rays, i.e.~null geodesics,
propagate at $45$ degrees in the $T,R$ plane, just as they did
in the original $t,r$ coordinates.
An analysis of the asymptotic behaviour of geodesics leads to the following 
results.
Future-directed timelike geodesics approach the point $(T,R)=(\pi,0)$,
which is therefore called \emph{future timelike infinity} $i^+$.
Similarly, past-directed timelike geodesics approach $(T,R)=(-\pi,0)$,
\emph{past timelike infinity} $i^-$.
Future-directed null geodesics approach the surface $T+R=\pi$,
\emph{future null infinity} $\mathrsfs{I}^+$ (``Scri$+$'').
Past-directed null geodesics approach $T-R=-\pi$, \emph{past null infinity}
$\mathrsfs{I}^-$.
Finally, spacelike geodesics approach $(T,R)=(0,\pi)$, \emph{spacelike
infinity} $i^0$.
Note that the conformal factor $\Omega$ in \eref{e:confmink} vanishes 
at $\mathrsfs{I}^\pm$.
We refer the reader to \cite{FrauendienerLRR} for an in-depth discussion
of conformal infinity.

Similar Penrose diagrams can be drawn for other spacetimes.
New features can arise, e.g.~singularities and event horizons in black hole
spacetimes.
For our purposes at this point, we are mainly interested in the asymptotic
region, in particular spacelike infinity and null infinity,
which is common to all asymptotically flat spacetimes.
Hence Minkowski spacetime will serve us as a representative example of an
asymptotically flat spacetime.


\subsection{Initial-boundary evolution}
\label{s:outer_ibvp}

\begin{figure}
  \centering
  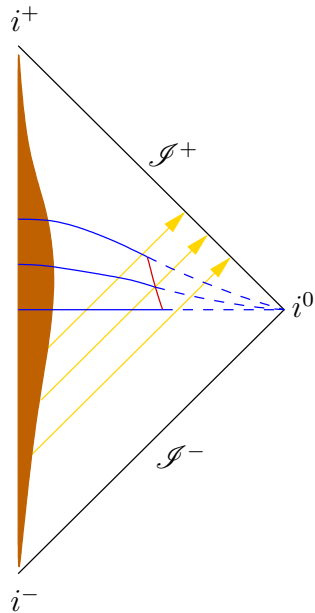
  \caption{\label{f:cauchy} 
  Cauchy evolution (blue lines) with artificial timelike boundary (red line).
  Shown is the Penrose diagram of Minkowski spacetime with a
  source (brown region) of radiation (yellow arrows).
  }
\end{figure}

The standard method for numerical evolutions of asymptotically flat spacetimes
is to foliate spacetime by spacelike hypersurfaces all approaching spacelike
infinity, drawn in blue in figure \ref{f:cauchy}, with initial data specified
on an initial slice.
Consider a sequence of signals propagating at the speed of light,
symbolised by the diagonal yellow lines in figure \ref{f:cauchy}.
Since all spatial slices approach $i^0$, these signals can never leave
the slices. 
Suppose we wanted to compactify the slices by mapping $i^0$ to a finite
spatial coordinate location.
Then an outgoing wave would appear increasingly ``blue-shifted'' (i.e.
with decreasing wavelength) with respect to the compactified coordinates,
and would ultimately fail to be resolved on the numerical grid.
Thus compactifying towards spacelike infinity is normally not a good idea.
(In \cite{Rinne2007} we assess the numerical performance of this approach,
among others.)

For these reasons one usually truncates the spatial slices at a finite 
distance.
This introduces an artificial timelike boundary, the red line in
figure \ref{f:cauchy}.
Boundary conditions must be imposed there so as to obtain a well-posed
initial-boundary value problem.
These boundary condtions are not arbitrary because the constraint equations
must hold on each individual slice.
Furthermore, ideally one would like the solution on the truncated domain to be
identical with the solution on the unbounded domain.
Spurious reflections of gravitational radiation should be avoided.
Such boundary conditions are called \emph{transparent} or \emph{absorbing}.
The first part of this thesis will be devoted to the analysis and numerical 
implementation of these questions, and will be summarised in section
\ref{s:ibvp} below.
For a comprehensive review article of this field of research 
see \cite{SarbachLRR}.

There is a fundamental problem with this approach:
in general relativity, gravitational radiation is only well defined
at future null infinity $\mathrsfs{I}^+$.
This is the result of the seminal work by Bondi, Sachs and coworkers in a 
series of papers from the 1960s \cite{Bondi1962}.
At a finite distance a ``local flux of gravitational radiation'' cannot be 
defined in the full nonlinear theory.
This is only meaningful if one linearises about a given background spacetime,
e.g.~Minkowski or more generally, Schwarzschild or Kerr spacetime.
Any absorbing boundary conditions imposed at a finite distance can therefore
only be approximate.


\subsection{Cauchy-perturbative and Cauchy-characteristic matching}
\label{s:matching}

One approach is to match the fully nonlinear evolution in the interior
to an outer module that solves the \emph{linearised} Einstein equations.
Gauge-invariant treatments of gravitational perturbations exist that require 
the solution of a scalar master equation, one for each pair $(\ell,m)$ with 
respect to a spherical harmonic expansion of the gravitational field.
These scalars are functions of $t$ and $r$ only so it is relatively inexpensive
computationally to move the outer boundary to a very large distance.
Some more details of this method are discussed in section \ref{s:absorbing}.
It should be stressed that the linearised equations are still solved on
Cauchy slices approaching spacelike infinity $i^0$.

\begin{figure}
  \centering
  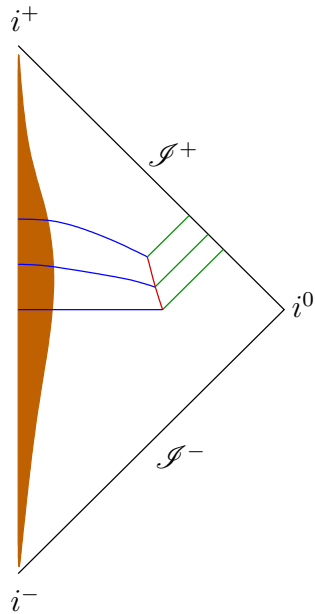
  \caption{\label{f:ccm} 
  Cauchy-characteristic matching. An inner Cauchy foliation (blue) is matched
  to an outer characteristic foliation (green).
  }
\end{figure}

A different approach is to attach to the truncated spacelike foliation
a \emph{characteristic} foliation extending to future null infinity
$\mathrsfs{I}^+$.
This is represented by the green lines in figure \ref{f:ccm}.
The ``blue-shift problem'' mentioned above does not apply to these 
null slices and hence it is straightforward to compactify them.
The difficult part of this method is the matching that needs to be done
at the artificial boundary.
So far this has been successfully implemented for \emph{a posteriori}
characteristic extraction, whereby one first carries out a Cauchy evolution
with boundary and then, in a post-processing step, reads out boundary data
for the subsequent characteristic evolution.
For this to work reliably, one needs to make sure that the artificial boundary
is placed sufficiently far out so that any inaccuracies emanating from it do 
not reach the extraction surface, which is rather wasteful.
So far the ultimate task of doing the matching ``on the fly'' while the
Cauchy evolution is still running has not been fully accomplished.
We refer to \cite{WinicourLRR} for a review of the 
Cauchy-characteristic matching approach.

The reader might wonder why one does not get rid of the spatial foliation
altogether and extend the characteristic slices all the way to the centre.
The reason is that null geodesic congruences, to which these slices are tied,
are generally ill behaved in strong-field regions:
they tend to form \emph{caustics}, which lead to coordinate singularities.
This caveat does not apply to situations with a high degree of symmetry,
e.g.~spherical symmetry, where characteristic evolution has indeed been
successfully used since the early days of numerical relativity.


\subsection{Hyperboloidal evolution}
\label{s:outer_scri}

\begin{figure}
  \centering
  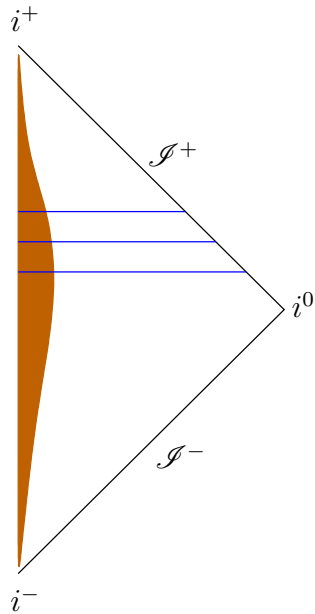
  \caption{\label{f:hyp} 
  Hyperboloidal evolution.
  }
\end{figure}

Yet another approach, taken in the second part of this thesis, is to foliate
spacetime by \emph{hyperboloidal} surfaces (figure \ref{f:hyp}).
These are spacelike but approach future null infinity rather than spacelike
infinity.
An example are the standard hyperboloids in Minkowski spacetime,
\[
  t = \sqrt{r^2 + \left( \frac{3}{K} \right)^2},
\]
where the constant $K$ turns out to be the mean curvature of the slices.
Such constant mean curvature surfaces can be constructed in more general 
spacetimes, and will be used in the second part of this thesis.
However other choices of hyperboloidal surfaces are possible.

The hyperboloidal initial value problem consists in specifying initial data 
on an initial hyperboloidal surface and evolving them to the future.
Note that hyperboloidal surfaces are only partial (future) Cauchy surfaces.

We will follow Penrose's idea and work with a conformally related metric
in a compactified coordinate system.
Unfortunately, the Einstein equations as such are not conformally invariant,
and as a result develop terms that are formally singular at $\scri$.
Dealing with these terms is the main challenge in \cite{Moncrief2009}.


\section{Cauchy evolution with artificial timelike boundary}
\label{s:ibvp}

This section summarises my work on initial-boundary value problems
for the Einstein equations, represented by the three papers 
\cite{Rinne2006}--\cite{Rinne2008b} in the first part of this thesis.

My interest in this topic arose during my time as a postdoc in the Caltech
group, who had just developed a first-order reduction \cite{Lindblom2006}
of the generalised harmonic formulation of the Einstein equations 
(section \ref{s:gh}).
They had proposed on physical grounds a set of boundary conditions that seemed
to work well in numerical simulations, and they were now interested in
proving that these boundary conditions actually rendered the initial-boundary
value problem well posed.


\subsection{Well posedness \cite{Rinne2006}}
\label{s:wp}

The generalised harmonic formulation is convenient from a mathematical point
of view because it is essentially a system of nonlinear wave equations,
and the initial-boundary value problem for such equations is relatively well
understood.
However several complications arise in the Einstein case.

For simplicity, let us consider the scalar wave equation (with a source $F$),
\[
  u_{tt} = u_{xx} + u_{yy} + u_{zz} + F
\]
on the half-space
\[
  x \geqslant 0, \quad -\infty < y < \infty, \quad -\infty < z < \infty
\]
with boundary conditions
\[
  \label{e:1storderbcs}
  \alpha u_t = u_x + \beta_1 u_y + \beta_2 u_z + \alpha q \quad \mathrm{at}\;x=0,
\]
where $\alpha>0$ is a constant and $q$ are boundary data.
The initial data are
\[
  u = f_1, \quad u_t = f_2 \quad \mathrm{at} \; t=0.
\]
One should think of $u$ as representing the individual components of the metric
in the generalised harmonic formulation of the Einstein equations.

For $\beta_1 = \beta_2 = 0$ the boundary conditions are maximally dissipative.
Defining the energy
\[
  E := \norms{u_t} + \norms{u_x} + \norms{u_y} + \norms{u_z}, 
\]
it is straightforward to obtain an estimate of the form
\[
  \label{e:strongwp}
  \fl \int_0^t  \norms{\mathbf{u}(s)} \rmd s
  + \int_0^t \normbs{\mathbf{u}(s)} \rmd s \
  \; \leqslant \; K_T \left( \norms{\mathbf{f}} + \int_0^t \norms{F(s)} \rmd s
      + \int_0^t \normbs{q(s)} \rmd s \right)
\]
for every finite time interval $0\leqslant t \leqslant T$ with a constant
$K_T$ that is independent of $F$, $f_1$, $f_2$ and $q$.
Here $\norm{\cdot}$ and $\normb{\cdot}$ denote the 
$L_2$ norms over the half-space and boundary, respectively, and
we have defined the vectors $\mathbf{u} := (u, u_t, u_x, u_y, u_z)$ and
$\mathbf{f} := (f_1, f_2, f_{1x}, f_{1y}, f_{1z})$.
The initial-boundary value problem is said to be \emph{strongly well posed}.

As already mentioned in section \ref{s:outer_ibvp}, boundary conditions for 
Einstein's equations must be compatible with the constraint equations on the 
$t=\const$ hypersurfaces.
The constraints satisfy a nonlinear wave equation of their 
own \eref{e:ghconstrevol}.
The simplest constraint-preserving boundary condition one could imagine is 
\[
  \label{e:dirichletcpbc}
  \mathcal{C}_a \hateq 0,
\]
where $\hateq$ denotes equality at the boundary.
This condition is of first order w.r.t.~derivatives of the metric, i.e.~of the
form \eref{e:1storderbcs}, but unfortunately with $\beta_1, \beta_2 \neq 0$,
i.e.~not maximally dissipative.
Later Kreiss and collaborators managed to prove strong well posedness 
for a set of boundary conditions including \eref{e:dirichletcpbc}
using energy methods with a non-standard choice of energy 
norm \cite{Kreiss2007}.

Still, the boundary conditions \eref{e:1storderbcs} are too restrictive in many
respects.
The constraint-preserving boundary conditions \eref{e:dirichletcpbc} are
a Dirichlet condition for a wave equation \eref{e:ghconstrevol}.
Consequently, any constraint violations generated in the interior will be 
reflected off the boundary.
Better behaved boundary conditions can be obtained by requiring the incoming
characteristic fields of \eref{e:ghconstrevol} to vanish at the boundary
so that the constraint violations will leave the domain.
However, this will involve first derivatives of the $\mathcal{C}_a$ and hence
second derivatives of the metric.
More seriously, absorbing boundary conditions will also involve second 
(or higher) derivatives of the metric.
This is because gravitational radiation is encoded in the Weyl tensor
$C_{abcd}$ (the tracefree part of the Riemann curvature tensor), which contains 
second derivatives of the metric.
In \cite{Rinne2006} we use as a ``physical'' boundary condition the vanishing
of a particular projection of the Weyl tensor, the Newman-Penrose scalar 
\[
  \label{e:Psi0}
  \Psi_0 = -C_{abcd} l^a m^b l^c m^d.
\]
Here the vectors on the right-hand side are part of a Newman-Penrose tetrad 
$(l^a, k^a, m^a, \bar m^a)$,
where $l^a$ and $k^a$ are outgoing and ingoing real null vectors satisfying
$l^a k_a = -1$, $m^a$ is a complex spatial null vector orthogonal to
$l^a$ and $k^a$, and $\bar m^a$ is its complex conjugate, with 
$m^a \bar m_a = 1$.
$\Psi_0$ can be regarded as an approximation to the incoming gravitational 
radiation.

For boundary conditions of higher derivative order than \eref{e:1storderbcs},
the energy method can no longer be applied.
Instead, pseudo-differential techniques can be used.
For the time being we assume the source terms $F$ and initial data $f_1$, $f_2$ 
vanish.
The idea is to perform a Fourier-Laplace transform and write the solution as
a superposition of modes
\[
  \label{e:fltransform}
  u(t,x,y,z) = \tilde u(x) \exp [st + \rmi (\omega_y y + \omega_z z)]
\]
with $s\in \mathbb{C}$ and $\omega_y, \omega_z \in \mathbb{R}$.
Suppose the homogeneous problem with vanishing boundary data ($q=0$)
admits a solution with $\Re s > 0$.
Then we obtain another solution by multiplying the exponent in 
\eref{e:fltransform} with any real number.
Hence the initial-boundary value problem cannot be well posed because the
growth of the solution cannot be controlled.

As reviewed in \cite{Rinne2006}, this condition amounts to showing that a 
certain complex determinant does not have any zeros $s$ with $\Re s > 0$,
the \emph{determinant condition}.
What remains to be shown is that for the inhomogeneous problem, the solution 
can be bounded in terms of the boundary data.
This turns out to be possible only if the zeros of the determinant have 
strictly negative real part, the \emph{Kreiss condition}.
If it holds then one obtains an estimate 
\[
  \int_0^t \norms{\mathbf{u}(s)} \rmd s \leqslant K_T \int_0^t \normbs{q(s)} 
\rmd s
\]
and the system is said to be \emph{boundary stable}.
The main result of \cite{Rinne2006} is that this condition holds for the given 
first-order reduction of the generalised harmonic Einstein equations and the
given boundary conditions.

A stronger estimate that includes the source terms $F$ on the right-hand side 
(cf.~\eref{e:strongwp}) is referred to as \emph{strong well posedness in the 
generalised sense}.
In addition to boundary stability this requires the construction of a 
symmetriser \cite{Kreiss2006}.
For the first-order reduction of the generalised harmonic Einstein equations
used in \cite{Rinne2006} it was not clear how to construct such a symmetriser.
What we do show though is that the Kreiss condition rules out
so-called weak instabilities with polynomial time dependence.

Later in \cite{Ruiz2007} strong well posedness in the generalised sense was 
proved for the original second-order form of the equations, avoiding 
complications arising from the first-order reduction.
From the theory of pseudo-differential operators it follows that strong well 
posedness in the generalised sense carries over to systems with variable
coefficients and quasi-linear systems such as the Einstein equations.

Lacking a full proof of strong well posedness, we perform numerical 
experiments in order to probe the stability of the system.
The numerical implementation uses pseudo-spectral methods as described in
section \ref{s:ps}.
The boundary conditions are implemented via a projection method,
which modifies the evolution equations at the boundary by eliminating
(derivatives of) the incoming fields using the boundary conditions. 

We perform \emph{robust stability tests}, whereby small random noise is 
injected in the initial data and source terms.
The background solution is taken to be either Minkowski spacetime on a spatial 
domain with topology $T^2 \times \mathbb{R}$ or Schwarzschild spacetime on
$S^2 \times \mathbb{R}$.
These experiments show no signs of instabilities and strongly support the 
claim that the system is well posed.
The expected instability for a deliberately chosen set of ill-posed 
boundary conditions is also reproduced.


\subsection{Numerical comparisons \cite{Rinne2007}}

Having constructed a set of stable (and most likely well-posed) boundary
conditions for the Einstein equations in generalised harmonic gauge,
the question arises how well these boundary conditions perform numerically
compared to other choices.
Perfect boundary conditions would produce a solution on the truncated domain
that agrees with the solution on the unbounded domain restricted to the
truncated region.
We can use this principle in order to assess the boundary conditions in the
following way.
First we compute a \emph{reference solution} on a very large domain.
Because of the finite speed of propagation for hyperbolic PDEs, we can choose
the boundary to be sufficiently far out so that any inaccuracies emanating
from it remain out of causal contact with the interior region where comparisons
will be made.
Next we perform an evolution with the same initial data on a domain that is 
truncated at a much smaller distance, where the boundary conditions are
imposed that are to be assessed.
Finally we compare the solution on the truncated domain with the reference 
solution.

The test problem chosen in \cite{Rinne2007} is a Schwarzschild black hole with
an outgoing gravitational wave perturbation.
The background spacetime is written in Kerr-Schild coordinates,
which penetrate the event horizon at $r=2M$. 
We can remove the interior of the black hole from the computational domain
by placing an \emph{excision} boundary just inside the event horizon.
At this interior boundary all characteristics leave the domain so that no
boundary conditions are required.
The gravitational wave perturbation is taken to be an exact solution 
of the linearised (about flat space) Einstein equations with quadrupolar
($\ell=2$) angular dependence \cite{Teukolsky1982}.
The wave is taken to be outgoing initially, with a Gaussian profile.
Of course the constraints must be solved in order to obtain a valid set of
initial data for the Einstein equations.

The numerical implementation uses the same pseudo-spectral methods as
in \cite{Rinne2006} and as described above in section \ref{s:ps}.

Once the wave reaches the outer boundary, the imperfect boundary conditions
will generate reflections, which propagate into the interior.
In order to assess the amount of reflections, we evaluate the following 
quantities.
\begin{enumerate}
  \item The difference $\Delta \mathcal{U}$ between the test solution and the 
    reference solution of all components of the metric and their first 
    derivatives, in a suitable norm (see \cite{Rinne2007} for details).
    It should be stressed that $\Delta \mathcal{U}$ is coordinate dependent so 
    it will measure how well the solutions agree 
    \emph{in the given coordinates}.
    While ``gauge reflections'' have no physical meaning, they do matter from
    a numerical point of view as one does not want to waste resolution
    on short-wavelength features that merely correspond to a coordinate
    transformation.
  \item Violations of the constraints $\mathcal{C}$, again in a suitable norm.
    This quantity tests how well the boundary conditions preserve the
    constraints.
  \item The difference between the test solution and the reference solution
    of the outgoing gravitational radiation as measured by the Newman-Penrose
    scalar (cf. \eref{e:Psi0})
    \[
      \Psi_4 = -C_{abcd} k^a \bar m^b k^c \bar m^d.
    \]
    This quantity is computed on a sphere close to the outer boundary
    of the truncated domain (a procedure often referred to as 
    \emph{wave extraction}) and compared to the reference resolution.
    From a physical point of view it is important to understand how the accuracy
    of the extracted waveform is affected by the choice of boundary conditions.
\end{enumerate}

The benchmark set of boundary conditions used in \cite{Rinne2007} are the 
boundary conditions constructed and analysed in \cite{Rinne2006}, 
with one small modification:
for the components of the metric that can be loosely identified as the
``gauge degrees of freedom'', a slightly different boundary condition is
used that differs from the original one only in a lower-order term.
With this extra term the gauge boundary condition is exactly absorbing
for a spherical ($\ell=0$) gauge wave.
This small modification is found to lead to a substantial reduction of the 
coordinate-dependent difference $\Delta \mathcal{U}$, whereas the constraints
$\mathcal{C}$ and physical radiation $\Delta \Psi_4$ are of course unaffected.

In the following we summarise the various alternate boundary conditions that
are investigated in \cite{Rinne2007}, along with their numerical performance.
\begin{enumerate}
  \item \emph{Freezing the incoming fields.} 
    In this approach the time derivatives of all the incoming fields 
    are required to vanish at the outer boundary.
    While these boundary conditions render the initial-boundary problem
    well posed, they are neither constraint preserving nor absorbing.
    For increasing numerical resolution the quantity $\mathcal{C}$ is seen
    to converge to a nonzero function.
    This demonstrates that one does in fact not obtain a solution to
    Einstein's equations with these simple-minded boundary conditions.
  \item \emph{Sommerfeld boundary conditions.} 
    This type of condition is often used in numerical relativity simulations
    based on the BSSN system and corresponds to imposing
    \[
      (\partial_r + \partial_r + r^{-1}) (g_{ab} - \eta_{ab}) \hateq 0
    \]
    on all components of the metric at the outer boundary, where $\eta_{ab}$
    is the flat (Minkowski) metric.
    The numerical performance is similar to the boundary conditions described
    above, with slightly reduced constraint violations.
  \item \emph{Kreiss-Winicour boundary conditions.} 
    These conditions, proposed in \cite{Kreiss2006}, consist in requiring the 
    harmonic constraints to vanish at the boundary, 
    equation \eref{e:dirichletcpbc} above.
    We compute the remaining incoming characteristic fields from the 
    Schwarzschild background solution.
    Although we expected this condition to be more reflective for constraint
    violations, we do not find any indications for this numerically.
    Apparently the constraint damping terms in our formulation are very 
    effective in reducing any constraint violations before they reach the 
    boundary.
    However we do see larger errors in the physical quantities
    $\Psi_4$ than with our benchmark boundary conditions, which include
    the condition $\Psi_0 \hateq 0$.
  \item \emph{Spatial compactification.} This approach is not technically
    a boundary condition; instead we compactify the spatial domain towards
    spatial infinity (see the discussion at the beginning of section 
    \ref{s:outer_ibvp} above).
    A certain form of spectral filtering is applied in order to damp the
    outgoing waves as they become increasingly ``blue-shifted''.
    This turns out to work quite well as far as constraint violations are 
    concerned, however the errors in $\Psi_4$ are significantly larger than
    with our benchmark boundary conditions.
  \item \emph{Sponge layers.}
    This method, often used in the context of spectral methods, adds 
    artificial damping terms to the evolution equations that are only active
    in a region close to the outer boundary, schematically:
    \[
      \partial_t u = \ldots - \gamma(r) (u-u_0),
    \]
    where $u_0$ refers to the background solution and the function
    $\gamma(r)$ is non-negligible only close to the outer boundary.
    This method is found to lead to a small amount of constraint violations
    and to considerable errors in the outgoing radiation $\Psi_4$.
\end{enumerate}

In summary, our boundary conditions outperform all the alternate methods
considered here.
We can even compare the reflection coefficient $\Psi_0 / \Psi_4$ 
with the prediction from linearised theory and 
find good agreement with our simulations.


\subsection{Absorbing boundary conditions \cite{Rinne2008b}}
\label{s:absorbing}

The boundary conditions used in \cite{Rinne2006,Rinne2007} included a condition 
on the vanishing of the Newman-Penrose scalar $\Psi_0$, which can be regarded
as an approximation to the outgoing gravitational radiation.
Using this condition was found to significantly reduce spurious reflections
of gravitational radiation.
It turns out that one can do better: there is a hierarchy of absorbing 
boundary conditions of the form
\[
  \label{e:B-S}
  [r^2(\partial_t + \partial_r)]^{L-1} (r^5 \Psi_0) \hateq 0.
\]
Here $L$ refers to an expansion of the gravitational field in spherical
harmonics.
The boundary condition \eref{e:B-S} is perfectly absorbing for linearised
gravitational waves on a flat background spacetimes for all spherical harmonic
modes $\ell \leqslant L$.
For $L=1$ we recover our original condition $\Psi_0 \hateq 0$.

The boundary conditions \eref{e:B-S} were first suggested by Buchman and 
Sarbach \cite{Buchman2006}.
They considered the linearised Bianchi equations, which describe the 
propagation of gravitational radiation and in vacuum take the form 
\[
  \nabla^a C_{abcd} = 0, 
\]
where $C_{abcd}$ is the Weyl tensor.
By expanding the fields in spherical harmonics and constructing exact solutions
to the linearised equations, the conditions \eref{e:B-S} were designed to 
eliminate the ingoing solutions.
Later Buchman and Sarbach generalised their method to a Schwarzschild 
background \cite{Buchman2007}.

In \cite{Rinne2008b} we reformulate the boundary conditions in a way that is 
both conceptually more straightfoward and more amenable to numerical 
implementation. 
Gravitational perturbations can be described by the gauge-invariant 
Regge-Wheeler-Zerilli (RWZ) scalars $\Phi_{\ell m}^{(\pm)}$ (see \cite{Sarbach2001}
and references therein).
These are complex quantities, one for each spherical harmonic index $(\ell,m)$
and for two parities: even ($+$) and odd ($-$).
On a flat background, they obey the master equation
\[
  \label{e:RWZ}
  \left[ \partial_t^2 - \partial_r^2 + \frac{\ell(\ell+1)}{r^2} \right] 
  \Phi_{\ell m}^{(\pm)} = 0.
\]
This equation is known as the Euler-Poisson-Darboux equation; it is of course 
just the scalar wave equation in disguise.
The general outgoing and ingoing solutions have the form
\[
        \Phi^{(\pm) \, \mathrm{out}}_{\ell m}(t,r) 
        = \sum_{j=0}^{\ell} \frac{f^{(\pm)}_{j\ell m}(t-r)}{r^j}, \qquad
        \Phi^{(\pm) \, \mathrm{in}}_{\ell m}(t,r) 
        = \sum_{j=0}^{\ell} \frac{g^{(\pm)}_{j\ell m}(t+r)}{r^j}.
\]
The precise form of the functions $f^{(\pm)}_{j\ell m}$ and $g^{(\pm)}_{j\ell m}$
does not matter here.
The key observation is that 
\[
  \label{e:phibc}
  B_L \Phi_{\ell m}^{(\pm)\, \mathrm{out}} := 
  [r^2(\partial_t + \partial_r)]^{L+1} \Phi_{\ell m}^{(\pm)\, \mathrm{out}} = 0
\] 
provided that $L \geqslant \ell$.
Using $B_L \Phi_{\ell m}^{(\pm)} \hateq 0$ as a boundary condition will therefore 
eliminate the ingoing solutions for all $\ell \leqslant L$.
These are nothing but the well-known boundary conditions of Bayliss and
Turkel \cite{Bayliss1980} for the scalar wave equation.
It is straightforward to relate them to conditions on the Newman-Penrose
scalar $\Psi_0$ and recover \eref{e:B-S}.

Equation \eref{e:phibc} contains higher derivatives, which are difficult
to treat numerically.
In \cite{Rinne2008b} we address this by introducing a set of auxiliary variables so
that \eref{e:phibc} can be written as a system of ODEs intrinsic to the 
boundary.

So far we have only considered the RWZ equation
\eref{e:RWZ}.
What we would really like is a set of boundary conditions for the Einstein
equations, in the generalised harmonic formulation already used in the
previous work.
Our algorithm thus consists in three steps:
\begin{enumerate}
  \item extraction of the RWZ scalars from the spacetime metric at the
    boundary,
  \item evolution of the system of ODEs for the auxiliary variables 
    that implements the desired absorbing boundary condition,
  \item construction of boundary data for certain incoming characteristic fields
    of the Einstein equations from the auxiliary variables.
\end{enumerate}
In \cite{Rinne2008b} we describe each of these steps in detail.

Step (iii) can also be used as a recipe for Cauchy-perturbative matching
(section \ref{s:matching})
in the context of the generalised harmonic formulation of the Einstein 
equations, as we could equally well take the boundary data from an outer 
module that evolves the RWZ equations directly.

We also remark that strong well posedness in the generalised sense 
(see section \ref{s:ibvp}) was proved in \cite{Ruiz2007} for the original
second-order form of the Einstein equations in harmonic gauge with the
new higher-order absorbing boundary conditions as well.

From the numerical point of view, an expansion of the fields in spherical 
harmonics is required.
This fits well with our pseudo-spectral method, which already uses spherical
harmonics as the angular basis functions.
However some slightly intricate transformations between different 
representations of tensor spherical harmonics need to be carried out
(see the appendix of \cite{Rinne2008b}).

In order to test our numerical implementation, we evolve initial data 
corresponding to outgoing solutions of the linearised Einstein equations 
with fixed spherical harmonic dependence $(\ell, m)$. 
For $\ell=2$ these were derived in \cite{Teukolsky1982}.
In \cite{Rinne2008c} I constructed analogous solutions for arbitrary $\ell$.
We evolve these initial data on a truncated spherical domain using our new
absorbing boundary conditions.
During the evolution we extract the RWZ scalars at the boundary and compare 
with the analytical solutions.
Since we evolve the full nonlinear Einstein equations, whereas the analytical
solutions are only valid to linear order, we perform evolutions with
different amplitudes of the initial data and check that any quantities
that should vanish at the linear level decay (at least) quadratically
with amplitude.
Using this method we show for our numerical evolutions in \cite{Rinne2008b} that our 
boundary conditions $B_L$ are indeed perfectly absorbing for all 
$\ell \leqslant L$.

While the boundary conditions do not eliminate incoming modes with $\ell > L$,
they reduce their amplitude significantly.
We compute the expected reflection coefficient analytically in linearised
theory and find good agreement with our numerical evolutions.
For instance, the $\ell=3$ incoming mode is suppressed by a factor of about 
$100$ when the $L=2$ absorbing boundary condition is used as compared with
$L=1$, which corresponds to the old $\Psi_0 \hateq 0$ 
condition\footnote{Here we have taken the radius of the outer boundary to be 
twice the wavelength.}.
This demonstrates the dramatic improvement achieved by these higher-order 
absorbing boundary conditions.


\section{Hyperboloidal evolution to future null infinity}
\label{s:scri}

Much progress has been made with initial-boundary value problems for the 
Einstein equations: well-posed formulations have been derived, particularly
in the context of generalised harmonic gauge, and improved absorbing boundary
conditions have been constructed and implemented.
The fundamental problem remains however that in the full nonlinear theory
of general relativity, boundary conditions imposed at a finite distance can 
never be perfectly transparent in the sense that the solution on the truncated
domain agrees with the solution on the unbounded domain.
The absorbing boundary conditions considered in \cite{Rinne2008b}
rely on the validity of the linear approximation about a given background 
spacetime (Minkowski in our case).

For this reason I became interested in hyperboloidal evolution, which aims to
place the outer boundary of the computational domain at future null infinity 
$\scri$, the only physically meaningful (conformal) boundary of spacetime.
This is the topic of the second part of this 
thesis \cite{Moncrief2009}--\cite{Rinne2013}.


\subsection{Regularity at future null infinity \cite{Moncrief2009}}

Most approaches to hyperboloidal evolution are based on Penrose's idea of 
a conformal transformation of the spacetime metric combined with a 
compactifying coordinate transformation, as described in 
section \ref{s:confinf} above.
Unfortunately the Ricci tensor is not conformally invariant and as a result
the Einstein equations contain inverse powers of the conformal factor,
which are singular at $\scri$.

In the early 1980s Friedrich \cite{Friedrich1983a} developed an elegant
solution to this problem by constructing a symmetric hyperbolic system of
PDEs that contained the Einstein equations but also evolution equations for
the Weyl curvature.
Remarkably, his equations are regular everywhere, including at $\scri$.
They are however rather complicated, which may explain why they have not made
their way into mainstream numerical relativity, despite a burst of activity 
in the late 1990s (see \cite{Husa2003} for a review).
Recently \cite{Beyer2012}
there has been a renewed numerical interest in these equations, 
especially concerning an extension \cite{Friedrich1998}
of Friedrich's original formulation that is able to address the intricate 
issues that arise where null infinity meets spacelike infinity.

Here we follow a different approach, proposed by Moncrief, that aims to 
tackle the (formally) singular terms in the Einstein equations directly.
We wanted to develop a system that is simpler
than Friedrich's regular conformal field equations and more similar to other
formulations already used by the numerical relativity community.
We work with an ADM-like formulation with elliptic gauge conditions:
constant mean curvature slicing and spatially harmonic coordinates, as
described in section \ref{s:adm} above.
In the spatially compact case the Cauchy problem for
these equations was shown to be well posed \cite{Andersson2003};
therefore we expect this formulation to be well behaved in our case as well,
although a formal proof of well posedness of the hyperboloidal initial value
problem with conformal boundary at $\scri$ is still lacking.

As expected we find that both the constraints and the evolution equations
contain terms involving inverse powers of the conformal factor $\Omega$,
which become singular at $\scri$.
This is not so much of a concern for the constraint equations, 
as one can always multiply the entire equation by a suitably high power
of $\Omega$ before solving it,
but for the evolution equations it seems at first sight that the
right-hand sides are singular so that stable evolution near $\scri$ cannot be
expected.
However in \cite{Moncrief2009} we show that the formally singular terms can actually
be evaluated explicitly at $\scri$ in a completely regular way provided
the constraints hold.

On a given hyperboloidal slice, we expand all the fields in finite Taylor 
series in $r$ near $\scri$, in adapted coordinates so that $\scri$ corresponds 
to an $r=\const$ surface.
Thanks to the degeneracy of the elliptic constraint equations at $\scri$,
we are able to evaluate the first few radial derivatives of the fields
at $\scri$ by inserting the Taylor expansions into the constraints.
More precisely, we obtain the first three radial derivatives of $\Omega$,
the zeroth and first radial derivative of the components $\pi^{\tr \, ri}$ of the
ADM momentum (directly related to the tracefree part of the extrinsic 
curvature),
and the first two radial derivatives of the conformal lapse 
function $\tilde \alpha = \Omega\alpha$.

With this information we are able to evaluate the formally singular terms
in the evolution equation for $\pi^{\tr \, ij}$ and show they are regular
at $\scri$, subject to one additional condition:
the vanishing of the shear of the null geodesic congruence that forms $\scri$.
This condition had already been found in \cite{Andersson1992}.
In \cite{Moncrief2009} we show in addition that it is preserved under the time evolution 
in the sense that if the shear vanishes at one instant of time, then its time 
derivative vanishes as well.

It is important to note that we only use \emph{finite} Taylor series at $\scri$.
We do not assume that the fields are smooth there.
The constraint equations give us just enough information about the first few 
derivatives of the fields at $\scri$ so that we can evaluate
the formally singular terms in the evolution equations.
In fact, it appears that in general, the Taylor expansion already breaks down 
at the next order and a polylogarithmic term needs to be 
included \cite{Bardeen2012}.
It could be that this is an artefact of CMC slicing.
Whether the polylogarithmic terms can be avoided in a different slicing
is an interesting open question.

There is a different, more straightforward way of deriving regular evolution
equations at $\scri$ by assuming that the conformal Weyl tensor vanishes
at $\scri$.
This \emph{Penrose regularity} implicitly assumes though that the conformal
metric is $C^3$ up to the boundary, a slightly stronger requirement than 
what we needed for our original analysis.
This different regularisation technique is also explored in \cite{Moncrief2009}.


\subsection{Axisymmetric reduction and numerical implementation 
  \cite{Rinne2010}}

In this section we describe the first successful numerical implementation
of the formulation developed in \cite{Moncrief2009}.
Since our regularity analysis at $\scri$ relied crucially on the satisfaction
of the constraint equations, we expect having to solve the constraints
explicitly at each timestep (constrained evolution).
This is computationally expensive and hence in this first application we
assume that spacetime is axisymmetric.
This reduces the number of effective spatial dimensions from three to two.
Unlike spherical symmetry, it is still compatible with gravitational radiation,
and we expect all the difficulties in the non-symmetric case already to be
present in axisymmetry as far as numerical stability at $\scri$ is concerned.
I had already developed a constrained axisymmetric evolution 
scheme on maximal Cauchy slices with timelike 
boundary \cite{RinnePhD,Rinne2008a} and hence it was obvious to try and adapt 
it to CMC slices with conformal boundary at $\scri$.

Spherical polar coordinates $(t,r,\theta,\phi)$ are used so that the Killing 
vector is $\partial/\partial\phi$, which in addition is assumed to be 
hypersurface orthogonal.
The spatial gauge condition used here differs from the spatial harmonic gauge
of \cite{Moncrief2009}. 
The conformal spatial metric $\tilde \gamma_{ij}$, which is related to the 
physical spatial metric via $\gamma_{ij} = \Omega^{-2} \tilde \gamma_{ij}$,
is taken to have the form
\[
  \label{e:quasiisotropic}
  \tilde \gamma = \rme^{2\eta\sin\theta} (\rmd r^2 + r^2 \rmd \theta^2)
  + r^2 \sin^2\theta \, \rmd\phi^2.
\]
This is known as \emph{quasi-isotropic gauge} and is chosen here because it
reduces the degrees of freedom in the conformal spatial metric to just one
function $\eta(t,r,\theta)$.
Preservation of this gauge condition in time yields a system of elliptic 
equations for the shift vector $\beta^i$.
In addition we need to solve the CMC slicing condition for the conformal lapse 
$\tilde \alpha$ and
the Hamiltonian constraint for the conformal factor $\Omega$.
There are evolution equations for the function $\eta$ and for the three
components of the tracefree part of the extrinsic curvature.

Even though the spatial gauge condition is different, the regularity analysis
at $\scri$ carries through as in \cite{Moncrief2009} and we obtain manifestly regular
forms of the evolution equations at $\scri$.
We have experimented with two slightly different versions, one derived
directly from the constraint equations using Taylor expansions,
the other by assuming the somewhat stronger Penrose regularity mentioned
in the previous subsection.
Numerically both appear to work equally well.

The numerical implementation is based on the finite-difference technique
(section \ref{s:fd}) with fourth-order accurate finite-difference operators.
The outermost radial gridpoint is placed right at $\scri$.
Here the regularised versions of the evolution equations are used, with
one-sided finite differences.
Already one further grid point in we have no choice but to use the full,
formally singular evolution equations.
Remarkably, this appears to be stable, provided the constraints are solved
at each substep of the Runge-Kutta time integration scheme.
We provide a heuristic explanation for the success of the method by observing
that the evolution equations contain terms that tend to push the solution 
towards the values dictated by the regularity conditions.

Some care needs to be taken when solving the elliptic equations using multigrid.
Since the equations degenerate at $\scri$, it is not surprising that a 
straightforward pointwise Gauss-Seidel relaxation fails to converge.
Instead, we use a radial line relaxation (with a direct one-dimensional solver)
and then perform Gauss-Seidel iterations in the angular direction.

As a first test problem we consider a Schwarzschild black hole.
The metric on CMC slices is known in closed form; we just need to compactify the
radial coordinate, which requires the numerical solution of one ODE.
An inner excision boundary is placed just inside the event horizon.
We are able to evolve initial data taken from this metric for times 
$t \sim 1000 M$ ($M$ being the black hole mass) without any signs of 
instability and with the expected fourth-order convergence as the numerical
resolution is increased.

Next we include a gravitational wave perturbation by specifying free initial 
data for the function $\eta$ in \eref{e:quasiisotropic}, which vanishes for the
unperturbed Schwarzschild spacetime.
We can read out the gravitational radiation at $\scri$ by computing the
gauge-invariant Bondi news function \cite{Bondi1962}, which can be computed
directly from the conformal spacetime Ricci tensor,
\[
  N = \bar m^a \bar m^b \tilde R_{ab},
\]
where the Newman-Penrose tetrad used must have the property that 
$m^a$ is tangential to $\scri$, i.e. $m^a \partial_a \Omega = 0$.
We observe the expected quasi-normal mode radiation generated by the perturbed
black hole (which essentially acts as a damped harmonic oscillator):
\[
  N_\ell \propto \rme^{-\kappa_\ell t} \sin (\omega_l t + \phi_\ell),
\]
where $\ell$ refers to the index of an expansion in spherical harmonics.
For the small perturbation we use ($\sim 10^{-4}$), 
the values of $\kappa_\ell$ and $\omega_\ell$ fitted from our numerical 
evolution are in good agreement with the semi-analytic results from linear 
perturbation theory.
At later times, when the quasi-normal mode radiation has decayed, one
expects a power-law \emph{tail} (often referred to as 
\emph{Price's law} \cite{Price1972})
\[
  N_\ell \propto t^{-p_\ell}.
\]
At the numerical resolutions that we are able to afford, we cannot see
this tail yet---as runs with two different resolutions demonstrate,
the solution has not converged yet.
The algorithm will need to be speeded up in order to study these 
subtle phenomena.


\subsection{Including matter; numerical evolutions in spherical 
   symmetry \cite{Rinne2013}}

Resolving late-time power-law tails of gravitational and matter fields on
black hole backgrounds is a very demanding problem.
With the current axisymmetric implementation of our hyperboloidal evolution
scheme we were unable to provide sufficiently high resolution.
In order to test if our method is capable to study tails in principle,
we decided to take one step back and impose spherical symmetry.

Due to Birkhoff's theorem, spherically symmetric vacuum spacetimes are
necessarily static: they are isometric to the Schwarzschild solution.
Thus in order to have non-trivial dynamics in spherical symmetry, matter
needs to be included.
How to deal with matter in the context of hyperboloidal evolution based on
conformal compactification is an interesting problem in its own right,
and so we investigated this quite generally, without any spacetime symmetries
at first.

We need to impose the condition that the energy-momentum tensor be
tracefree,
\[
  \label{e:Ttf}
  g^{ab} T_{ab} = 0.
\]
Under this assumption the energy-momentum conservation equations, which 
constitute the evolution equations for the matter fields, are conformally
invariant: 
if we define a conformally related energy-momentum tensor 
$\tilde T_{ab} := \Omega^{-2} T_{ab}$ then standard energy-momentum conservation
$g^{ab} \nabla_a T_{bc} = 0$ implies that
\[
  \label{e:ConfEMC}
  \tilde g^{ab} \tilde \nabla_a \tilde T_{bc} = 0,
\]
where $\tilde \nabla$ is the connection compatible with the conformal spacetime 
metric $\tilde g_{ab}$.
Without the condition \eref{e:Ttf}, the equations \eref{e:ConfEMC} contain
an additional term that is singular at $\scri$.
Condition \eref{e:Ttf} is generally satisfied for ``radiative'' forms of 
matter such as a (conformally coupled) massless scalar field, Maxwell or 
Yang-Mills fields.
It is not satisfied e.g.~for a general perfect fluid.
However if the support of the matter remains compact during the evolution 
then one needs not to worry about the singular terms at $\scri$.

We work out the matter evolution equations and matter source terms in the 
Einstein equations explicitly for two examples: 
a conformally coupled scalar field and Yang-Mills fields.

The Einstein-scalar field equations arise from varying the action
\[
  S = \int ( \tfrac{1}{2\kappa} R - \tfrac{1}{2} g^{ab} \phi_{,a} \phi_{,b}
  - \tfrac{1}{12} R \phi^2 ) \sqrt{-g} \, \rmd^4 x.
\]
The last term is referred to as \emph{conformal coupling} and leads to a 
conformally invariant evolution equation for the scalar field $\phi$:
\[
  \Box \phi - \tfrac{1}{6} R \phi = 0 \quad \Leftrightarrow \quad
  \tilde \Box \tilde \phi - \tfrac{1}{6} \tilde R \tilde \phi = 0,
\]
where $\Box$ is the d'Alembert operator, as above a tilde refers to the 
conformal spacetime metric, and we have introduced a rescaled scalar field
$\tilde \phi := \Omega^{-1} \phi$.

Yang-Mills theory can be regarded as a nonlinear generalisation of 
electromagnetism to non-abelian gauge groups.
Its fundamental field is a vector potential or connection $A_a^{(\alpha)}$.
The upper index refers to the gauge group, which we will take to be 
SU(2), so Greek indices range over $1,2,3$ here.
The associated field strength tensor is
\[
  \label{e:ymf}
  F_{ab}^{(\alpha)} = \partial_a A_b^{(\alpha)} - \partial_b A_a^{(\alpha)}
  + f^{\alpha\beta\gamma} A_a^{(\beta)} A_b^{(\gamma)}.
\]
Note the last term, which is absent in electromagnetism.
The symbol $f^{\alpha\beta\gamma} = g \, [\alpha\beta\gamma]$ is totally 
antisymmetric, where $[123] := +1$ and $g$ is a dimensionful coupling constant.
Repeated Greek indices are summed over.
The Yang-Mills field equations are given by
\[
  \nabla_a F^{ab\,(\alpha)} + f^{\alpha\beta\gamma} A_a^{(\beta)} F^{ab\,(\gamma)} = 0.
\]
They have the convenient property to be conformally invariant and hence
we may adorn all quantities in the above equations with tildes
and work directly in the conformal spacetime.
When performing the $3+1$ decomposition, the Yang-Mills equations split into
a constraint and an evolution equation.

After this general discussion and examples of matter models we reduce the
equations to spherical symmetry.
Isotropic spatial coordinates are chosen so that the conformal spatial metric
is flat.
The tracefree part of the extrinsic curvature has only one free component
in this case.
Unlike in \cite{Rinne2010}, we solve the momentum constraint for it,
rather than its formally singular evolution equation.
(This is only possible in spherical symmetry.)

While the reduction to spherical symmetry is straightforward for the Einstein
and scalar field equations, it is not so obvious for the Yang-Mills fields.
The most general spherically symmetric (conformal) Yang-Mills connection has 
the form
\[
  \label{e:yma}
  \tilde A^{i(\alpha)} = [\alpha ij] x^j F + (x^\alpha x^i - r^2 \delta^{\alpha i})H
  + \delta^{\alpha i} L, \quad \tilde A_0^{(\alpha)} = G x^\alpha,
\]
where $F,H,L$ and $G$ are functions of $t$ and $r$ only.
In most previous numerical studies only the potential $F$ was included;
we present for the first time evolutions with fully general spherically 
symmetric Yang-Mills fields.

Our numerical method is very similar to the one of \cite{Rinne2010}.
Since there is only one spatial dimension now, the constraint equations
are ODEs, which we solve using a direct band-diagonal solver combined with an
outer Newton-Raphson iteration to address the nonlinearity.

The initial data are chosen to be either Minkowski or Schwarzschild spacetime
(in CMC slicing) with an approximately ingoing matter perturbation (scalar
field or Yang-Mills).
On the flat background we are able to take the amplitude to be large enough
so that a black hole forms during the evolution, and to continue the 
evolution after excising its interior.

With the increased numerical resolution that is possible in the spherically
symmetric case, we can now see the tails and measure their decay exponents.
The results are in good agreement with previous numerical work.
This includes two studies that also used hyperboloidal 
evolution \cite{Puerrer2005,Puerrer2009}, however in coordinates that 
are not horizon-penetrating so that gravitational collapse could not be
studied.

A general property of power-law tails is that the decay at $\scri$ is slower
than at a finite distance.
(It would be impossible to see this with a code based on Cauchy evolution
with artificial timelike boundary!)
This causes the solution to resemble a ``boundary layer'' at late times
and the runtime of the simulation at fixed resolution is limited (though
sufficient in our case to obtain reliable results).

One feature we find that does not seem to have been noticed before is that
in the Yang-Mills case, the electric field (a component of the field strength
tensor \eref{e:ymf}) has a slower decay rate at $\scri$ ($\sim t^{-1}$) than
the connection ($\sim t^{-2}$).
Furthermore, for the general spherically symmetric connection \eref{e:yma}
we find some interesting gauge dynamics:
while all components of the energy-momentum tensor decay so that a vacuum
solution is approached, the components of the connection approach a constant
or even time-periodic solution in some cases.
We explain this behaviour by deriving the most general form of the 
spherically symmetric vacuum solutions to the Einstein-Yang-Mills system.


\section{Conclusions and outlook}
\label{s:concl}

This thesis is concerned with analytical and numerical approaches to treating
the far field of asymptotically flat spacetimes satisfying the Einstein
equations.
We focus on two different approaches: Cauchy evolution with artificial 
timelike boundary (part 1) and hyperboloidal evolution to future null infinity
(part 2).

In the first part, we prove a necessary condition (boundary stability)
for well posedness of the initial-boundary value problem for a first-order
reduction of the Einstein equations in generalised harmonic gauge with
constraint-preserving boundary conditions.
These include a condition on the Weyl tensor component 
$\Psi_0$, which can be regarded as a first approximation to the incoming
gravitational radiation.
Numerical stability tests further demonstrate the robustness of the boundary
conditions.
Next we assess the numerical performance of various other boundary conditions
and alternate approaches such as compactification to spacelike
infinity or sponge layers by comparing the solution on the truncated domain
with a reference solution on a much larger domain.
In all cases our boundary conditions are found to be superior.
Finally we formulate and implement a hierarchy of higher-order absorbing
boundary conditions that improve on the original $\Psi_0 \hateq 0$ condition.
Our approach is based on the Regge-Wheeler-Zerilli scalars, and we show how
it can be interfaced with the generalised harmonic formulation of the Einstein
equations.

In the second part, we work with a constrained ADM-like formulation of the
Einstein equations on constant mean curvature slices extending to future
null infinity.
Upon a conformal transformation of the metric, the Einstein equations develop
terms that are formally singular at future null infinity $\scri$.
However, we show explicitly how these terms can be evaluated at $\scri$
in a completely regular way.
Based on this idea we present a first numerical implementation for 
vacuum axisymmetric spacetimes.
Long-term stable evolutions of a gravitationally perturbed Schwarzschild black 
hole are obtained and the Bondi news function, which describes the outgoing 
gravitational radiation in a gauge-invariant way, is evaluated at $\scri$.
Finally we extend our formulation to include matter with trace-free 
energy-momentum tensor. 
Scalar and Yang-Mills fields are coupled to the Einstein equations and
evolved numerically in spherical symmetry.
This includes spacetimes that form a black hole from regular initial data.
We study the power-law decay (``tail'') of the matter fields at late times, 
both at $\scri$ and at a finite distance.

There are a number of ways in which the research presented in this thesis can 
be extended.
We discuss both parts separately.

Concerning Cauchy evolution with artifical boundary, it would of course be
nice to complete the proof of strong well posedness in the generalised sense
for the particular first-order reduction of the Einstein equations in 
generalised harmonic gauge and boundary conditions we used.
However, given that there is already a proof for the original second-order
system and that the boundary conditions appear to be very robust numerically,
there is currently not so much interest in this question.
Our implementation of absorbing boundary conditions could be generalised by
allowing for a Schwarzschild rather than flat background spacetime.
In general however, the numerical relativity community seems to be quite
happy with their current codes and seem to be reluctant to invest much
effort in improved boundary conditions.
This may well change once gravitational wave astronomy has advanced to a
stage that even more accurate simulations are required.

Certainly from the current point of view, hyperboloidal evolution appears to
be a much cleaner solution to the outer boundary problem.
Our axisymmetric numerical implementation demonstrates that stable numerical
evolutions based on our approach can be achieved,
however the code will need to be speeded up in order to be useful in 
practice, especially in the case without symmetries.
For instance, one could try to solve the constraints explicitly only in a 
neighbourhood of $\scri$ and use free evolution in the interior.
We also intend to generalise our formulation to more general gauge conditions,
as we do not believe the particular gauge we used (constant mean curvature
slicing and spatially harmonic coordinates) was essential for the regularity
analysis at $\scri$.
Hyperboloidal evolution should have interesting applications whenever
global properties of spacetime are to be investigated.
An example is cosmic censorship, as one can now check whether null
geodesics manage to escape to future null infinity.


\section*{Acknowledgments}

I would like to thank my Habilitation committee at FU Berlin 
(Ralf Kornhuber, Klaus Ecker, Konrad Polthier, Theodora Bourni, Florian 
Litzinger) and the referees for their time and effort, and for making me 
welcome at the Department.
Further thanks go to my colleagues at the Albert Einstein Institute and my
collaborators over the past years who contributed to this work.
Support through a Heisenberg Fellowship and grant RI 2246/2 of the German 
Research Foundation (DFG) is gratefully acknowledged.


\section*{References}

\providecommand{\newblock}{}

\end{document}